\documentclass{aa}  
\usepackage{graphicx}
\usepackage{txfonts}
\usepackage[colorlinks]{hyperref}
\usepackage{lineno}

\usepackage{subcaption}
\usepackage{float}

\begin{document}

\definecolor{forestgreen}{rgb}{0.10, 0.50, 0.10}
\newcommand{\peijin}[1]{{{}{#1}}}
\newcommand{\diana}[1]{\textbf{\textcolor{forestgreen}{#1}}}
\newcommand{\emilia}[1]{\textbf{\textcolor{forestgreen}{#1}}}

\title{Spatially resolved radio signatures of \\
electron beams in a coronal shock}

\titlerunning{Interferometry of solar herringbone burst}
\authorrunning{Zhang et. al.}

   \author{Peijin Zhang
          \inst{1}
          \and
          Diana Morosan \inst{1}\fnmsep\thanks{corresponding author}
          \and
          Anshu Kumari \inst{1,2}
          \and
          Emilia Kilpua \inst{1} 
          }

   \institute{Department of Physics, University of Helsinki, P.O. Box 64, FI-00014, Helsinki, Finland\\
              \email{peijin.zhang@helsinki.fi, diana.morosan@helsinki.fi, emilia.kilpua@helsinki.fi}
         \and
             Heliophysics, NASA Goddard Space Flight Center, Greenbelt, MD 20771, USA\\
             \email{anshu.kumari@nasa.gov}
             }

   \date{Received \today; accepted ***}

\abstract
{Type II radio bursts are solar radio burst associated with coronal shocks. Type II bursts usually exhibit fine structures in dynamic spectra that represent signatures of accelerated electron beams. So far, the sources of individual fine structures in type II bursts have not been spatially resolved in high-resolution low-frequency radio imaging.} 
{The objective of this study is to resolve the radio sources of the herringbone bursts found in type II solar radio bursts and investigate the properties of the acceleration regions in coronal shocks} 
{We use low-frequency interferometric imaging observations from the Low Frequency Array (LOFAR) to provide a spatially resolved analysis for three herringbone groups (marked as A, B, and C) in a type II radio burst that occurred on 2015 October 16th.}
{The herringbones in groups A and C have a undiversified frequency drift direction and propagation direction along frequency. They have similar value of frequency drift rates corresponding to that of type III bursts and previously studied herringbones. 
Group B has a more complex spatial distribution, with distinct sources separated by 50~arcsec and no clear spatial propagation with frequency. One of the herringbones in group B was found to have an exceptionally large frequency drift rate (373\,MHz/s).}
{The imaging spectroscopy features suggest that the studied herringbones originate from different processes.  The herringbone groups A and C most likely originate from single-direction beam electrons, while group B can be explained by counterstreaming beam electrons.} 

   \keywords{solar -- radio -- shock}

   \maketitle

\section{Introduction} \label{sec:intro}

Low-frequency solar radio emissions are dominated by solar radio bursts that are classified into five main types (Type I--V) based on their shape and characteristics in dynamic spectra. Type II solar radio bursts are characterized as slow frequency-drifting bursts towards lower frequencies and they often exhibit complex structures in the dynamic spectra \citep[e.g.,][]{wild1950typeII, Mann1995AAcharacteristicTypeii, Cairns2003SSRtypeII, ramesh2023ApJsolarcoronadens}. 
Statistical and independent studies have shown that type II solar radio bursts are in general associated with coronal mass ejections \citep[CMEs;][]{Subramanian2006AAstatistical,Gopalswamy2019catalog,chen2014ApJtypeiiCME,Majumdar2021SoPh, morosan2021, kumari2023type}. In rare cases, type II bursts have been reported to occur in the absence of a CME.  However,  in these cases, the presence of a coronal shock wave can be inferred from multi-wavelength observations \citep[e.g.,][]{su2015, morosan2023type}.

Type II bursts are interpreted as the signature of shock-accelerated electron beams generating Langmuir waves close to the local plasma frequency ($f_p$), and then converted into radio emission at fundamental (F) and harmonic (H) frequency of $f_p$ \citep{NelsonMelrose1985srph, Cairns2003SSRtypeII}. The volume emissivity of the plasma emission is strongly dependent on the plasma density distribution and the electron beams \citep{Schmidt2014JGRA,Cairns2003SSRtypeII, kumari2017b}. As the shock can create complex plasma distributions around it \citep{Hegedus2021ApJ}, type II solar radio bursts usually have complex fine structures. 
One common fine structure in type II radio bursts is the herringbone structure \citep[e.g.,][]{Cairns1987SoPh, carley2013, Diana2019NatAs, morosan2022}, which manifests as a series of parallel lines in the dynamic spectra resembling the pattern on a fish's backbone where its name derives. With the increasing resolution of the dynamic spectra provided by the recent radio instruments, a wide variety of fine structures in type II radio bursts have been reported \citep{Jasmina2020ApJfine}.

Imaging spectroscopy of type II radio bursts can be used to locate shock-accelerated electron beams \citep[e.g.,][]{Zucca2018AAShock,Diana2019NatAs}. With the new availability of low-frequency radio imaging, recent studies have provided improved multi-frequency images of type II bursts and associated fine structures \citep[e.g.,][]{Zucca2018AAShock, Diana2019NatAs, chr20, bhunia2022imagingspectroscopy}. For example, \citet{Zucca2018AAShock} used the tied-array beam observation mode \citep{mo14} with the Low-frequency array \citep{van2013lofar} for a type II solar radio burst. The observation is combined with an MHD model of the solar corona to obtain the 3D source location. The results of the above-stated study indicate that the type II burst is located at the CME flank where the shock proceeding direction is quasi-perpendicular to the background magnetic field. \cite{Diana2019NatAs} also used tied-array beam imaging observations with LOFAR to image herringbone bursts and identified the presence of shock-accelerated electron beams at multiple regions around the CME shock.
\cite{Maguire2021lofar} used LOFAR interferometry imaging combined with EUV observation on a type II radio burst, showing the radio source is located about 0.5 solar radii above a solar jet, which suggests the type II was generated by a jet-driven piston shock, instead of a CME-driven shock wave.
In another study with the Murchison Widefield Array (MWA), \cite{bhunia2022imagingspectroscopy} did radio imaging for a band-splitting type II event at multiple frequencies, and found that the radio source is located at multiple locations close to the shock.
Spatially resolved radiometry analysis is a powerful tool to analyze in detail the electron beam locations and propagation by imaging the fine structures in dynamic spectra.  

\begin{figure*}[h]
    \centering
    \includegraphics[width=0.75\paperwidth]{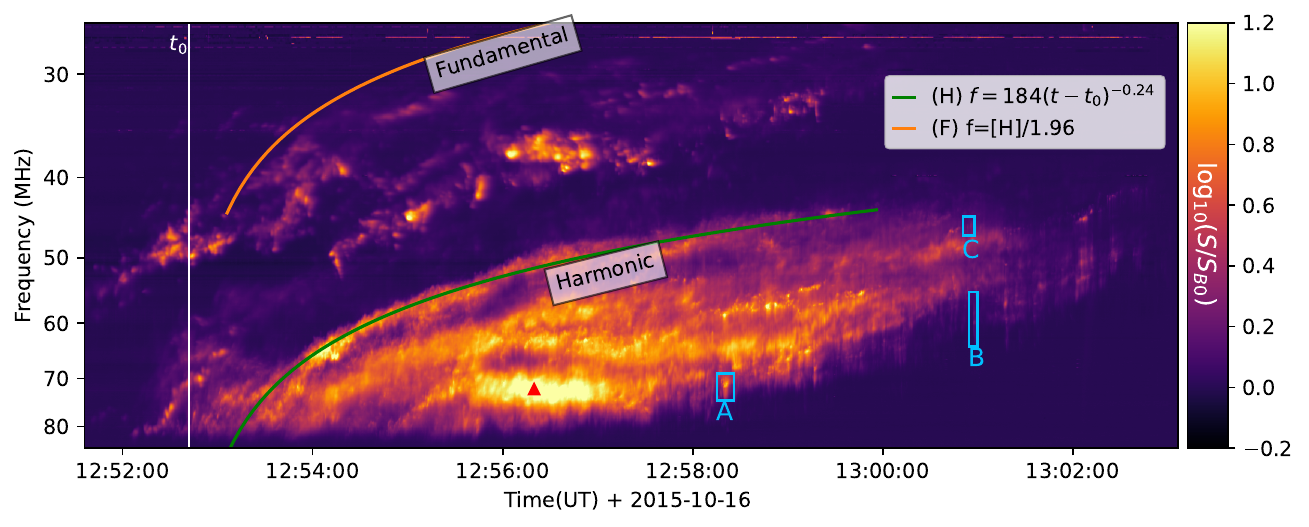}
    \caption{The dynamic spectrum of the type II radio burst, the flux is normalized according to the quiet time flux ($S_{B0}$), representing the relative flux in reference to the quiet Sun. 
    The fitted leading-edge of the Harmonic part of this type II is marked as green solid lines, the orange solid line indicates the leading-edge of the fundamental emission, derived from multiplying the radio of $\rm F/H=1/1.96$.
    Three herringbone groups are marked by blue boxes and labeled as A, B, and C.
    The brightest time-freq point is marked as red triangle.}
    \label{fig:overview}
\end{figure*}

In this paper, we present the first high \textbf{spatial}-resolution low-frequency study of the source of herringbone bursts. This paper is arranged as follows: in Section 2, we introduce the instrument and observation methods including the imaging spectroscopy method and the data reduction procedure. Section 3 presents the type II radio burst event in the dynamic spectrum. Section 4 presents the interferometry observation of the herringbone structures in the type II radio burst. In section 5, we analyzed the observation results and compared them with previous studies.
The overall conclusion is summarized in section 6.

\section{Observation and data reduction} \label{sec:floats}


We observed a type II radio burst with LOFAR's Low Band Antennas (LBAs) on 16 October 2015 in the frequency range 10--90~MHz. The observation mode used included imaging observations in interferometric mode and beamformed observations with one beam pointed at the Sun to produce a dynamic spectrum. The dynamic spectrum measures the total flux from the target (the Sun), and the time and frequency resolution of the dynamic spectrum is 10.4 ms and 12.5 kHz.
The dynamic spectrum is processed with \texttt{ConvRFI}  \citep{zhang2023rfi} to remove radio frequency interference. 
\peijin{For each individual herringbone lane structure, to get the frequency drift rate (FDR) df/dt, we perform a linear fit to the local maximum pixels in the herringbone lane.  The fitting target is the reciprocal of the FDR dt/df (as df/dt can approach $\pm$inf in this case). Uncertainty for each peak time pixel is half-cadence (5.2ms), which is taken into account to estimate the uncertainty.}

The interferometric observation used 23 core stations and 12 remote stations. These 35 stations form 561 cross-correlation baselines in total. Two sub-array pointings were arranged to target the Sun and Virgo-A (calibrator source).
For interferometric imaging, the integral time and frequency are 0.25 seconds and 195.3 kHz, respectively. The frequency channel spacing is, however, not uniform: in the range 20-58\,MHz it is 195.3 kHz and in the range 58-80\,MHz the spacing is 390.5 kHz. So far, this is the only LOFAR interferometric observation available with such a high-frequency resolution that allows us to study the evolution of low-bandwidth herringbones and other fine structures in type II bursts.

The type II burst on 16 October 2015, as shown in Fig (\ref{fig:overview}), is a fundamental-harmonic (F-H) pair event. The frequency ratio of the H- and F-components is 1.96 when comparing the bright part of each component. Band-splitting is visible in both F- and H- components. 
There are many complex structures in the harmonic lane, including herringbone bursts.
We picked a few relatively isolated herringbone bursts in this type II event, labeled as A B, and C. We picked the herringbone structures that were most isolated from the background emission of the type II burst so that there is less confusion on the radio source identification of herringbones.
The harmonic emission in the dynamic spectrum has a relatively distinct leading edge, we fitted the leading edge of the harmonic leading edge to $f\rm[MHz]\it=A (t-t_0)^{B}$ where $f$ is the frequency in [MHz], $t$ is the time offset in seconds, $t_0$ is the launching time of the shock inferred from the leading edge, the fitted value of $t_0$ is 16 October 2015 12:52:41.717 UT. We also perform source kinematics analysis using interferometric images.

\section{Results}

This type II event features many complex structures in the harmonic lane, including numerous herringbone bursts. The three groups of herringbones selected, A B, and C, are analyzed in detail below.

\subsection{Herringbone group A}

\begin{figure*}[ht]
    \centering
    \begin{subfigure}{0.39\paperwidth}
        \includegraphics[width=\textwidth]{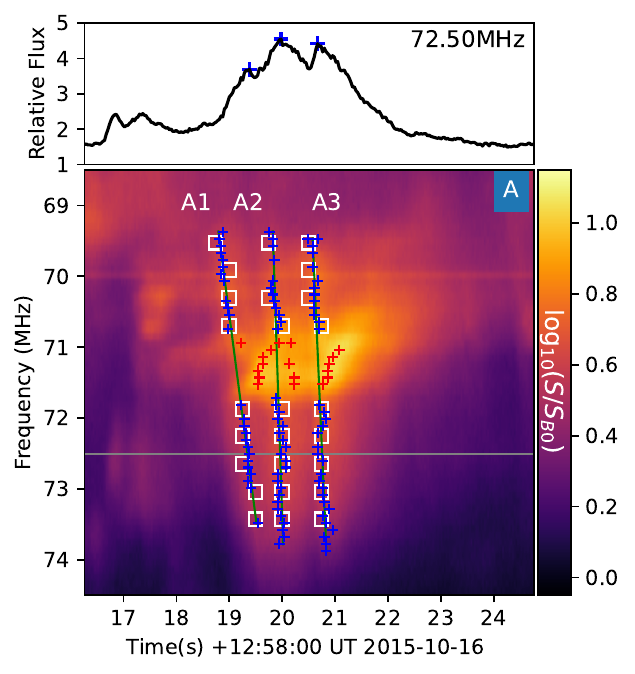}
        \caption{Dynamic spectrum of Herringbone group A, consisting of three individual herringbone structures (A1, A2, A3). The upper panel is the flux of 72.5MHz (marked as a gray solid line in the dynamic spectrum), the blue plus marks the local maximum points along the herringbone, red plus marks the local maximum on the overlapped structures.
        The flux is in the relative unit in reference to the quiet time flux ($S_{B0}$) before the burst time.
        The green line marks the frequency drift track of the herringbone.
        White rectangles marked the frequency and time integral span for the interferometric imaging.
        \peijin{The upper panel is the flux slice from 72.5MHz (gray line in lower panel)}}
        \label{fig:dsA}
    \end{subfigure}
    \hspace{5pt}
    \vspace{8pt}
    \begin{subfigure}{0.36\paperwidth}
        \includegraphics[width=\textwidth]{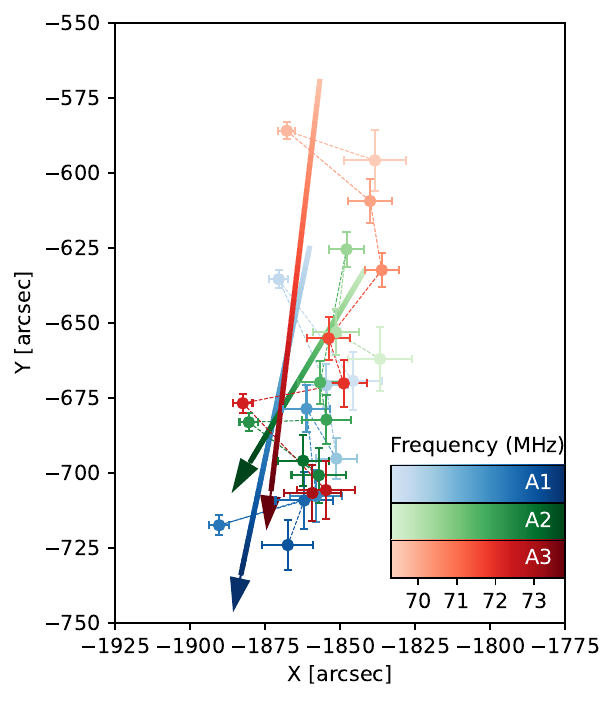}
        \caption{Source position of the herringbone structures in sub-event A, we used Gaussian fit to determine the source location, error-bar indicates the location uncertainty. Three herringbone components (A1, A2, and A3) are presented in Blue, Green, and Red. The brightness of the color represents the frequency. \peijin{The arrow indicates position variation track with frequency, the dashed lines connect to each point in order of frequency}.}
        \label{fig:eAfit}
    \end{subfigure}
    \begin{subfigure}{0.76\paperwidth}
        \includegraphics[width=\textwidth]{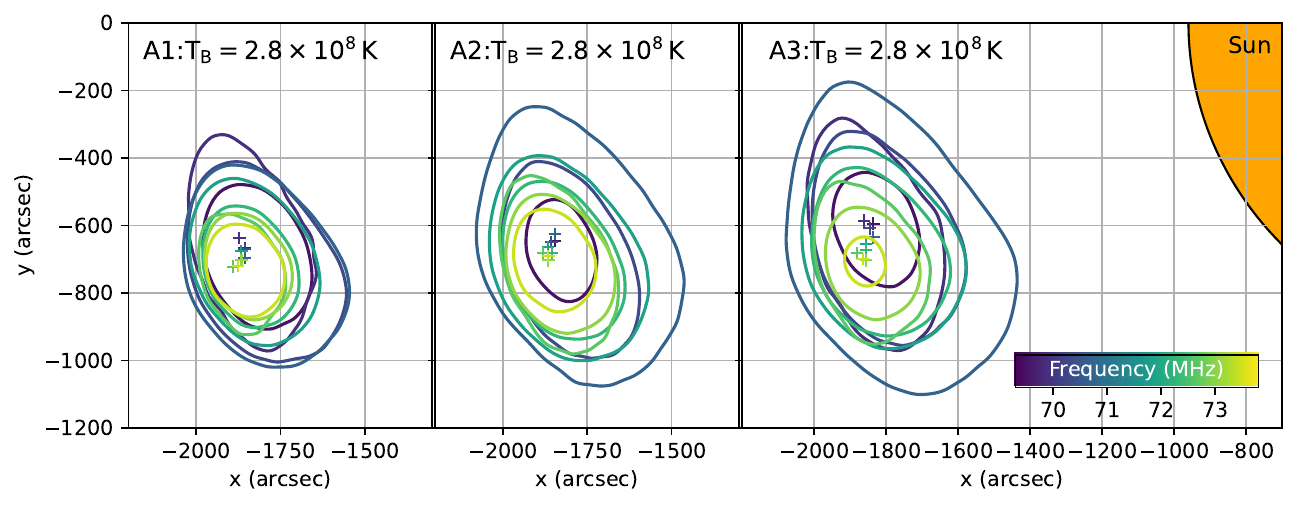}
        \caption{ Interferometric imaging of the herringbone at different frequencies, colored solid lines indicate the brightness temperature contour at 280MK, and the plus sign marks the peak location of the brightness temperature distribution.
        The time and frequency slot is marked in Fig (\ref{fig:dsA}), snapshot image in Fig (in Fig \ref{fig:imA1},\ref{fig:imA2},\ref{fig:imA3}).}
        \label{fig:eA_im}
    \end{subfigure}
    \caption{Spectral characteristics and spatial location of herringbone group A}
    \label{fig:A}
\end{figure*}

\begin{figure*}[ht]
    \centering
    \begin{subfigure}{0.40\paperwidth}
        \includegraphics[width=\textwidth]{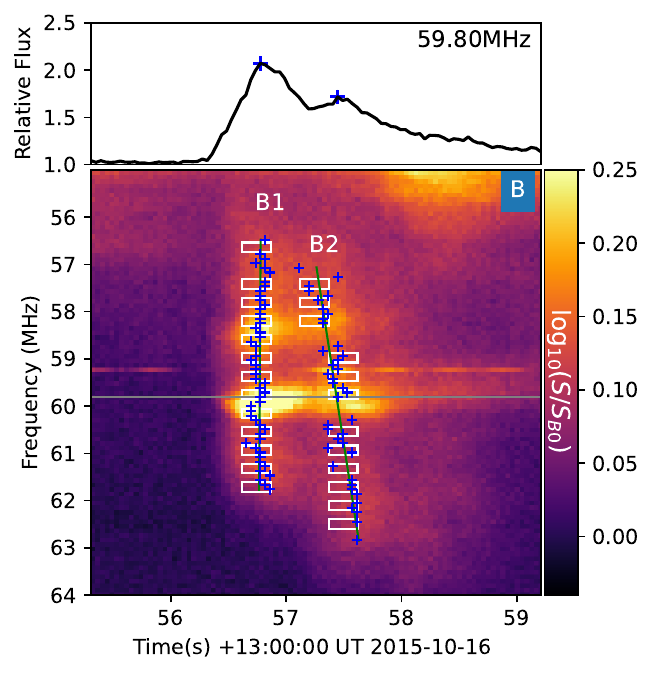}
        \caption{Dynamic spectrum of Herringbone group B, consisting of two individual herringbone structures (B1, B2). The upper panel is the flux of 59.9MHz (marked as a gray solid line in the dynamic spectrum), the blue plus marks the local maximum points along the herringbone, red plus marks the local maximum on the overlapped structures.
        The flux is in the relative unit in reference to the quiet time flux ($S_{B0}$) before the burst time.
        The green line marks the frequency drift track of the herringbone.
        White rectangles marked the frequency and time integral span for the interferometric imaging.}
        \label{fig:dsB}
    \end{subfigure}
    \hspace{5pt}
    \begin{subfigure}{0.35\paperwidth}
        \includegraphics[width=\textwidth]{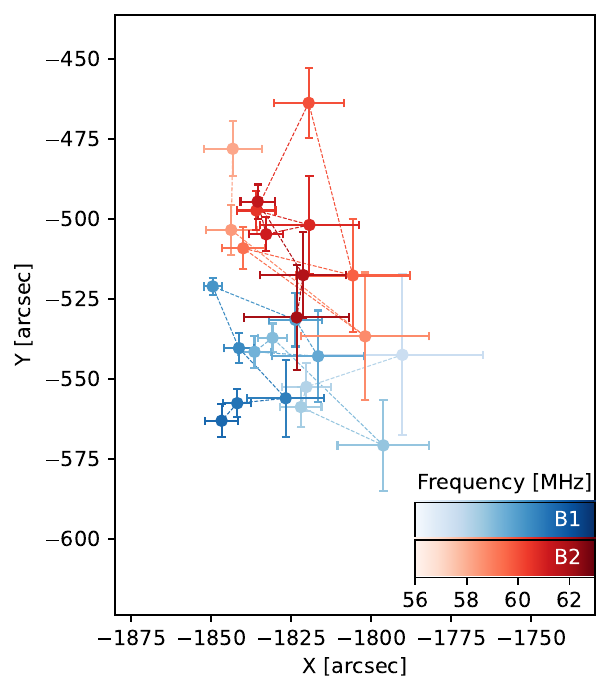}
        \caption{Source position of the herringbone structures in herringbone group B, we used Gaussian fit to determine the source location, error-bar indicates the location uncertainty. Three herringbone components (B1, and B2) are presented in Blue and Red. The brightness of the color represents the frequency. \peijin{ The dashed lines connect to each point in order of frequency,  the dashed lines connect to each point in order of frequency}} 
        \label{fig:imbfit}
    \end{subfigure}
    \hfill
    \begin{subfigure}{0.67\paperwidth}
        \includegraphics[width=\textwidth]{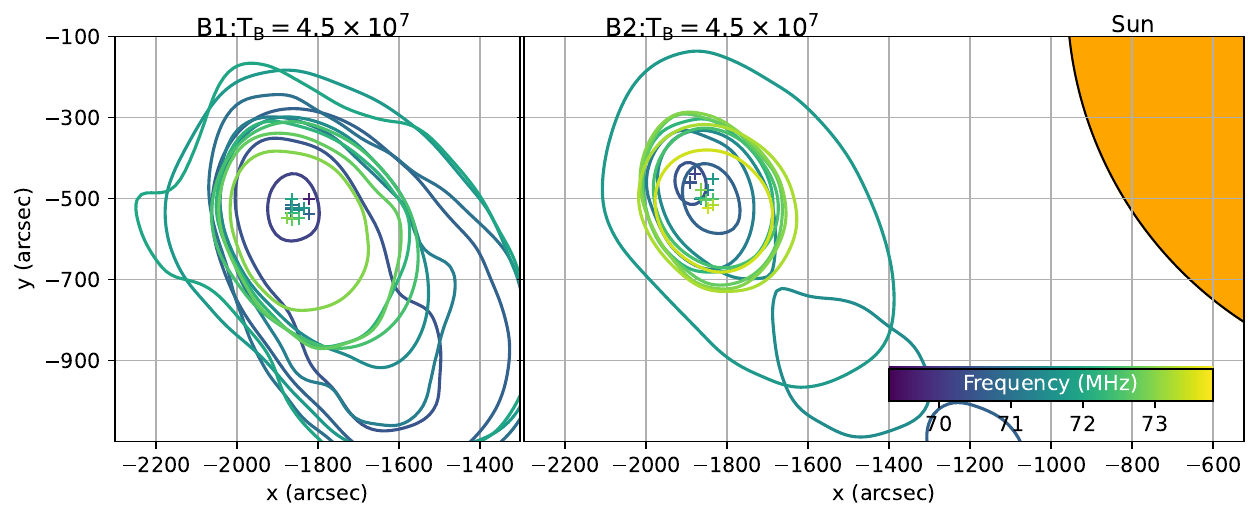}
        \caption{ Interferometric imaging of the herringbone at different frequencies, colored solid lines indicate the brightness temperature contour at 60MK, and the plus sign marks the peak location of the brightness temperature distribution.
        The time and frequency slot is marked in Fig (\ref{fig:dsB}), snapshot image in Fig (in Fig \ref{fig:imB1}, and \ref{fig:imB2}).}
        \label{fig:eB}
    \end{subfigure}
    \caption{Spectral characteristics and spatial location of herringbone group B}
    \label{fig:B}
\end{figure*}

\begin{figure*}[ht]
    \centering
    \begin{subfigure}{0.40\paperwidth}
        \includegraphics[width=\textwidth]{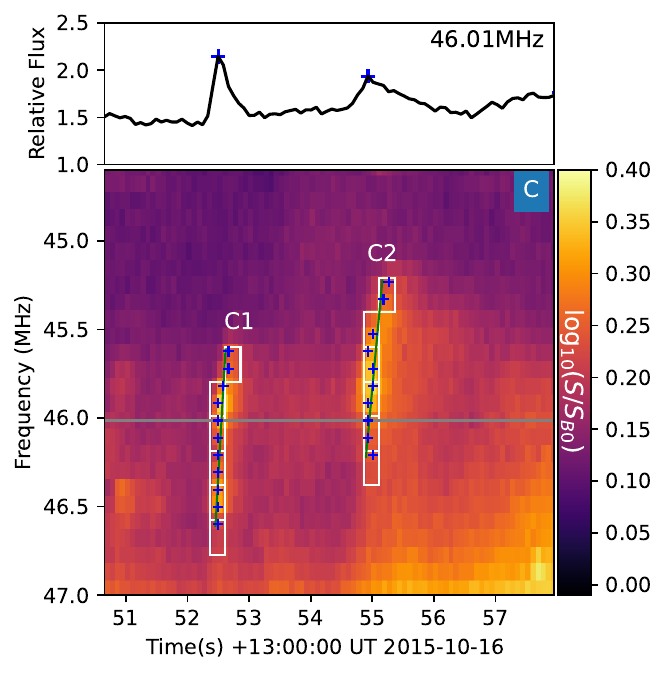}
        \caption{Dynamic spectrum of Herringbone group C, consisting of two individual herringbone structures (C1, C2). The upper panel is the flux of 46.0MHz (marked as a gray solid line in the dynamic spectrum), the blue plus marks the local maximum points along the herringbone, red plus marks the local maximum on the overlapped structures.
        The flux is in the relative unit in reference to the quiet time flux ($S_{B0}$) before the burst time.
        The green line marks the frequency drift track of the herringbone.
        White rectangles marked the frequency and time integral span for the interferometric imaging.}
        \label{fig:dsC}
    \end{subfigure}
    \hspace{5pt}
    \vspace{8pt}
    \begin{subfigure}{0.36\paperwidth}
        \includegraphics[width=\textwidth]{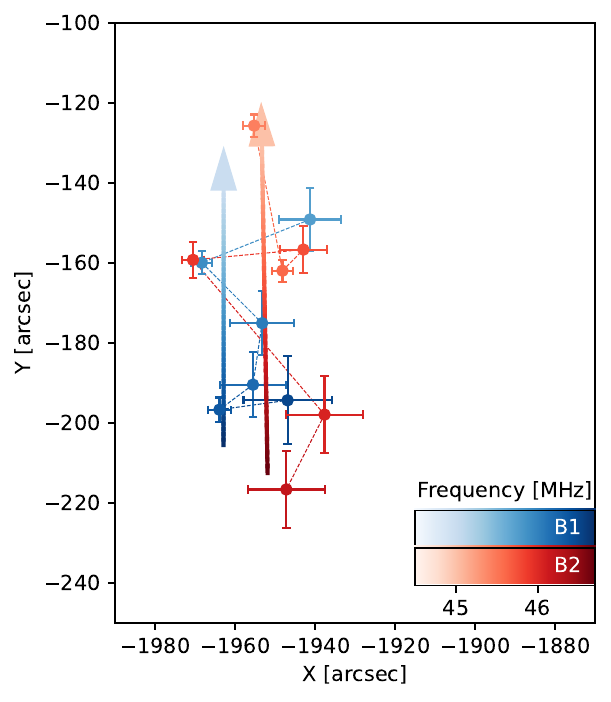}
        \caption{Source position of the herringbone structures in herringbone group C, we used Gaussian fit to determine the source location, error-bar indicates the location uncertainty. Three herringbone components (C1, and C2) are presented in Blue and Red. The brightness of the color represents the frequency. \peijin{The arrow indicates position variation track with frequency}.}
        \label{fig:Cfit}
    \end{subfigure}
    \begin{subfigure}{0.8\textwidth}
        \includegraphics[width=\textwidth]{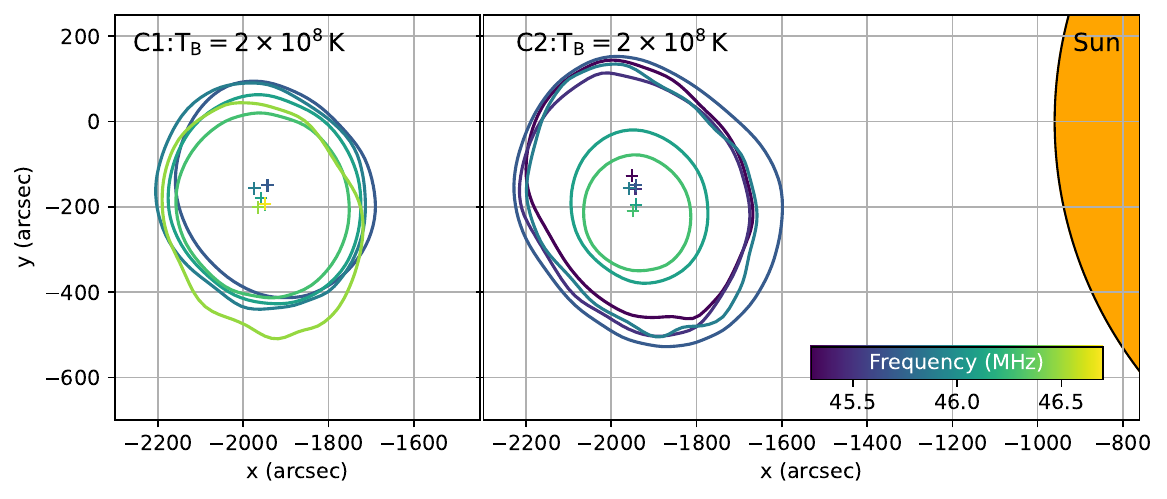}
        \caption{Interferometric imaging of the herringbone at different frequencies, colored solid lines indicate the brightness temperature contour at 200MK, and the plus sign marks the peak location of the brightness temperature distribution.
        The time and frequency slot is marked in Fig (\ref{fig:dsC}), snapshot image in Fig (in Fig \ref{fig:imC1}, and \ref{fig:imC2}).}
        \label{fig:eC}
    \end{subfigure}
    \caption{Spectral characteristics and spatial location of herringbone group C}
    \label{fig:C}
\end{figure*}

Sub-event A is a group of herringbones that occurs at 12:58:20~UT and around a frequency of 71~MHz. These herringbones have a reverse drift rate from low to high frequencies. The dynamic spectrum of this sub-event is shown in Fig. \ref{fig:dsA}. There are three herringbone components marked as A1, A2, and A3. We obtained the local maxima points for each frequency channel in the dynamic spectrum to locate the herringbones. The marked local maxima points present the track of the herringbone in the frequency range of 69.3-70.9 MHz and 71.5-73.5 MHz (shown as blue plus symbols in the lower panel of \ref{fig:dsA}).
There are also structures in the frequency range of 70.9-73.5 MHz \peijin{that do not follow the frequency drift lane of the herringbone structure. (local maxima marked as red symbols). Although the imaging of the excluded part (Fig \ref{fig:Ainout}) indicates that the source location is similar to the Herringbone Group A. To perform a more rigorous analysis of the Herringbone, the overlapped part is thus excluded from the following imaging analysis..}
By fitting the local maxima points of each herringbone component, we obtained the average frequency drift for each burst. 
\peijin{Fitted value of dt/df (1/FDR) for A1, A2 and A3 are $0.1746\pm0.0048 \rm (MHz/s)^{-1}$ , $0.044\pm0.0071 \rm(MHz/s)^{-1}$, and $0.0590\pm0.0068\rm (MHz/s)^{-1}$.
The corresponding frequency drift rates of each are 5.7MHz/s, 22.6MHz/s, and 16.9MHz/s, respectively.}

Imaging of this herringbone group is done inside the time-frequency blocks marked in Fig.~\ref{fig:dsA}, covering all local-maximum peak times for all available subbands.
The width of each subband and the integral time of each time slot for imaging is indicated by the height and width of each weight block in Fig \ref{fig:dsA}. The radio source shape and location of each herringbone component are shown in Fig. \ref{fig:eA_im}. The contour plot in this figure shows that the herringbones consist of a single source and have a simple geometrical shape. 


The radio source locations show a frequency dependence that is demonstrated by a 2D spline fit, as shown in Figure 
 \ref{fig:eAfit}. This outlines the frequency dispersion of herringbones. The herringbones trace a path from north to south for all three components (A1, A2, A3) and denote a common alignment of source location variation from high to low frequency. The starting and ending locations of the three tracks are separated by 20--50 arcsec. Considering the positive frequency drift rate, the temporal variation of the source is from north to south. Thus, the herringbones propagate southwards in the plane-of-sky.  Some projection effects may however be present.

\subsection{Herringbone group B}

Group B occurs at 13:00:57~UT and around a frequency of 60MHz. It consists of two herringbones, labeled B1 and B2, as shown in Fig. \ref{fig:dsB}. The two herringbones in this group are marked as blue plus symbols in the dynamic spectrum. By fitting the local maxima points of each herringbone component, we obtained the average frequency drift rate for each of these herringbones.
\peijin{Fitted value of dt/df (1/FDR) for B1, and B2 are $-0.0026\pm0.0043\rm(MHz/s)^{-1}$, and $0.0634\pm0.0071\rm(MHz/s)^{-1}$.
The corresponding frequency drift rates of each are  -373.0\,MHz/s and 15.7\,MHz/s, respectively.}
B2 has a reverse frequency drift like the herringbones in group A, however, B1 appears almost vertical in the dynamic spectrum, \peijin{fitted value of 1/FDR is close to 0, and the error range covers 0, indicating the frequency drift rate is either >598MHz/s or <-142MHz/s. The frequency drifting direction tends to be negative as the center value is -373.0MHz/s.}

Imaging of these herringbones shows complex spatial structures (see Fig. \ref{fig:eB}). Additional detailed images at different frequencies are shown in Figs.
\ref{fig:imB1} and \ref{fig:imB2}. In some of the frequency subbands covering the imaging observations, we observe multiple peaks in intensity (e.g. <~60~MHz in \ref{fig:imB2}) that are combined to form a larger elongated radio source in these images. There are also frequency subbands with less complex features in radio images (e.g. >~62~MHz in Fig. \ref{fig:imB2}. \peijin{To identify the herringbone source from multi-source images, we perform imaging to the adjacent time-freq point, as shown in Fig \ref{fig:inoutlabel}, inside herringbone structure, there are two sources (upper and lower), outside the herringbone, the upper source disappears. Also, }by comparing the multi-source images and the single-source images, the single sources appear to have well-aligned source sizes and positions with the upper west source in the multi-peak images. This source location also coincides with the time and frequency of the herringbones. Thus the source location and dynamic analysis in this section are focused on the upper west source from the multi-source images.

The source location of the herringbones in this group is shown as centroids in Fig. \ref{fig:imbfit}. The sources of the two components are not superposed as expected, instead, they are separated by 50\,arcsec. The track traced by the radio sources with frequency is also not spatially aligned, unlike the herringbones in the first group (A1--3).
The source locations of B1 and B2 do not exhibit any clear dispersion in position with frequency, as expected from plasma emission. B1 shows a trend from west to east as frequency decreases from high to low, and the source of B2 shows a trend from east to west as frequency decreases from high to low.

\subsection{Herringbone group C}

The third group of the herringbone structure occurs around a frequency of 46MHz and at 13:00:52\,UT. This group has two herringbones labeled C1 and C2 that are temporally separated by 2.5 seconds. These herringbones have a forward drift from high to low frequencies. By fitting the local maxima points of each herringbone component, we obtained the average frequency drift for each. The frequency drift rate of C1 and C2 is -6.1\,MHz/s and -3.7\,MHz/s, respectively.

Imaging of this herringbone group is done inside the time-frequency blocks marked in Fig. \ref{fig:dsC}. The radio source's shape and location of each component are shown in \ref{fig:eC}. The contour indicates that the radio source of this event is a single source and relatively simple in spatial structure.

The source centroids of this herringbone group are presented in Fig. \ref{fig:Cfit}. The frequency trend of the centroids is shown in a 2D spline fit. The track of these two components is spatially well aligned and shares the same position-frequency relation: from high to low frequency moving northwards, opposite to the herringbones in group A. This confirms the theory that opposite drifting herringbones also propagate in opposite directions, as expected from the plasma emission mechanism \citep{Melrose1980emission}. The electron beams producing the herringbones in groups A and C that move in opposite directions also confirm that bi-directional electron beams can escape the shock \citep{Zlobec1993}.

\begin{figure}[ht]
    \centering
    \includegraphics[width=0.42\paperwidth]{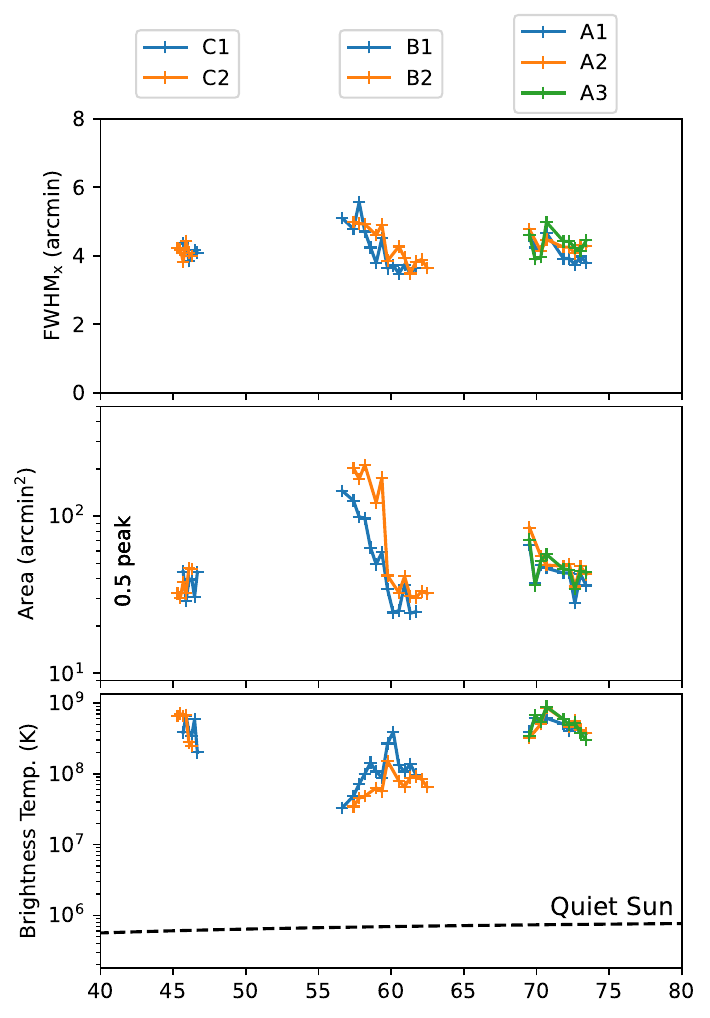}
    \caption{Source size and brightness temperature of the herringbone structure, area of the source is measured at half peak.}
    \label{fig:ABC}
\end{figure}

We also investigate some general properties of the herringbones in the three groups. Fig. \ref{fig:ABC} shows the brightness temperature and source size of these three herringbone groups.
The peak brightness temperature of these three groups ranges from $2\times10^7\,\rm K$ to $9\times10^8\,\rm K$. Group A has the highest brightness temperature, while Group B has the lowest brightness temperature. The size is measured \peijin{with 0.5peak area and $\rm FWHM_x$. The 0.5peak counts the area of the region within the contour of 0.5 times the peak $T_B$. The $\rm FWHM_x$ measures from the Full Width of Half Maximum (FWHM) in x direction, obtained with the following procedure: First, locate the upper source's coordinates. Then, slice in the x-direction from the peak, and obtain the FWHM from that slice. }
The $0.5\rm~peak$ area of group B has a significant variation at 59.5\,MHz. The area at 58\,MHz is $\sim$6-8 times larger than the area at 62\,MHz in the same group. In this group, a frequency of 59.5MHz marks the change between elongated and simple sources (see Figs. \ref{fig:imB1} and \ref{fig:imB2} where there are multiple sources and complex structures below 59.5\,MHz, and the source structure is relatively single and clear for the source above 59.5\,MHz).
The FWHM in the x-direction for the three herringbone groups are all close to 4\,arcmin. This indicates a source size significantly smaller than that of type III radio bursts \citep{Kontar2019ApJAnisotropicscattering}.

\begin{table*}[ht]
\centering
\begin{tabular}{lccccc}
\hline
 & \multicolumn{1}{l}{\begin{tabular}[c]{@{}l@{}}Freq Range\\ (MHz)\end{tabular}} & \multicolumn{1}{l}{\begin{tabular}[c]{@{}l@{}}Freq Drift Rate\\ (MHz/s)\end{tabular}} & \multicolumn{1}{l}{\begin{tabular}[c]{@{}l@{}}Max $T_b$ \\ ($\times 10^8$ K)\end{tabular}} & \multicolumn{1}{l}{\begin{tabular}[c]{@{}l@{}}Area of 0.5$\times$peak\\ ($\rm arcmin^2$)\end{tabular}} & \multicolumn{1}{l}{\begin{tabular}[c]{@{}l@{}}$\rm FWHM_x$\\ ($\rm arcmin$)\end{tabular}} \\
 \hline
A1 & {[}69.5, 73.5{]} & 5.7 & 6.2 & {[}28.81, 71.91{]} & 4.11 \\
A2 & {[}69.4, 72.7{]} & 22.6 & 8.7 & {[}36.53, 90.74{]} & 4.32 \\
A3 & {[}69.5, 73.5{]} & 16.9 & 8.9 & {[}35.47, 77.01{]} & 4.35 \\
B1 & {[}56.5, 61.8{]} & -373.0 & 4.1 & {[}25.24, 154.2{]} & 4.18 \\
B2 & {[}57.1, 62.8{]} & 15.7 & 1.6 & {[}31.51, 220.9{]} & 4.27 \\
C1 & {[}45.6, 46.6{]} & -6.1 & 6.6 & {[}30.81, 54.17{]} & 4.15 \\
C2 & {[}45.2, 46.2{]} & -3.7 & 7.1 & {[}32.38, 57.38{]} & 4.13 \\
\hline
\end{tabular}
\caption{Frequency range, frequency drift rate, and the brightness temperature of the source}
\label{tab:1}
\end{table*}

 Table \ref{tab:1} summarises the frequency range, frequency drift rate, maximum brightness temperature, area, and FWHM size in the x direction.
 The obsolete value frequency drift rate of most of the traces are within the range of $3-30 MHz/s$.
 Trace B1 has an exceptionally high frequency-drift-rate: -373.0\,MHz/s.
 There is no apparent correlation between the frequency drift rate, frequency range, and brightness temperature.

\section{Discussion}

Herringbone structures in type II solar radio bursts offer a unique diagnostic method to understand the small-scale processes of electron kinematics close to the shock. The characteristics of the radio emission, combined with models and assumptions, can provide information on how electron beams are accelerated and escape the shock. The following are several conclusions that can be deduced from our observations.

\subsection{Frequency drift rate}

Assuming that one herringbone is generated by a single-direction electron beam, the frequency drift rate of the herringbone corresponds to the speed of the beam. The method of deriving the beam speed from frequency drift rate is often used in type III radio burst analysis \citep[e.g.][]{zhang2018type}.
\cite{mel2005solar} reported 1\,MHz to 3\,MHz frequency drift rate in the range of 10\,MHz to 30\,MHz for both positive and revert drift.
\cite{carley2015low} reported $8.1^{+5.9}_{-6.1}$\,MHz for reverse drift and $-4.6^{+1.8}_{-3.9}$\,MHz for forward drifting herringbones in the range of 10$\sim$90\,MHz.
Based on previous observations \citep[e.g.][]{carley2015low, Cairns1987SoPh, mann2005electron}, the frequency drift direction stayed the same within the same group.

From our measurements of seven herringbones, the three herringbones in group A have positive drift, and the two in group C have negative drift. The drift rates of the herringbones from groups A and C are consistent with the results from \cite{carley2015low,mann2005electron}.
The two herringbones in group B, however, have opposite frequency drift directions and B1 has an exceptionally large absolute value of the frequency drift rate.
Using the coronal density model of \cite{saito77}, 373\,MHz/s corresponds to 5.92 times the speed of light, which is unphysical. There are two possible explanations for the exceptionally high frequency-drift-rate: (1) fast electron beams passing through a region with a drastic change in background plasma density; (2) multiple regions producing radio emission at different frequencies simultaneously. The second scenario can be achieved from multiple electron beams generating counterstreaming beam emission \citep{ganse2012emission}.

\subsection{Counterstreaming or single direction  beam electron}

\begin{figure}[ht]
    \centering
    \includegraphics[width=0.41\paperwidth]{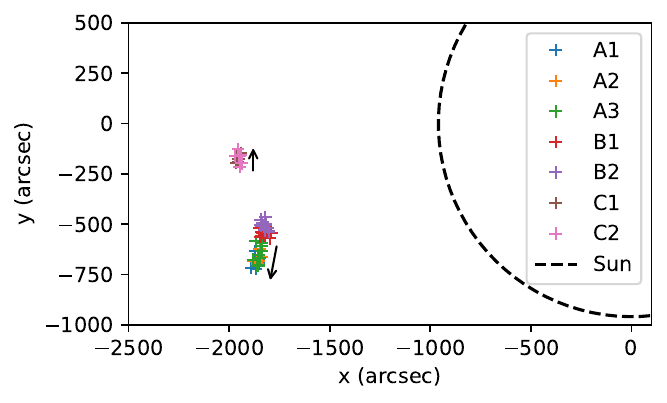}
    \includegraphics[width=0.39\paperwidth]{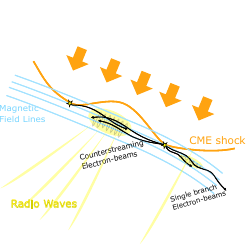}
    \caption{Combined plot of the coordinates of the sources of the three herringbone groups (upper panel) and Illustrative cartoon model depicting the radio emission source of the herringbone structures (lower panel).}
    \label{fig:cartoon}
\end{figure}

It is widely accepted that herringbones originate from electron beams that undergo acceleration by the CME shock \citep[e.g.][]{Cairns1987SoPh, Zlobec1993, Diana2019NatAs}. However, the specific mechanism behind this emission process remains uncertain with two possible ways of generating radio emission: (1) single-direction beam triggering Langmuir waves that are converted to radio waves through wave-wave interactions \citep{Melrose1980emission}, similar to Type III radio bursts and (2) counterstreaming electrons generate Langmuir waves, then the waves from the two beams interact and generate the radio waves \citep{knock2003type,ganse2012emission}. 
In the latter, using particle-in-cell simulations \citep{ganse2012emission}, radio emission can be detected in the case of counterstreaming beam electrons, while it is not detected in the single-direction electron beam scenario. This indicates that the requirement for generating radio waves is less strict for counterstreaming beam electrons. The possible emission mechanism is shown in the cartoon in Fig. \ref{fig:cartoon}, where a shock front (orange solid lines) propagates forward into quasi-perpendicular magnetic fields, creating multiple electron acceleration points (marked as the yellow stars).
The accelerated electrons can move along the magnetic field without encountering other electrons, namely as single-direction electron beams, shown in the lower-left corner of Fig. \ref{fig:cartoon}. Some electron beams can encounter another stream of electrons moving in the anti-parallel direction due to multiple shock ripple 
 acceleration and reflection \cite{knock2003type}, namely the counterstreaming region, shown in the center part of Fig. \ref{fig:cartoon}.

These two types of electron beam distributions (single direction and counterstream) can have different spectroscopic features. The multiple bursts in a herringbone group from a single-direction electron beam should have similar source locations and moving directions aligned with the propagation direction of the radio sources. The frequency drift rate would also be similar to type III bursts.
The herringbones generated by counterstreaming electrons would have arbitrary trajectories, as the encountering region could have different spatial distributions for each herringbone. The frequency drift rate is also possible to have extreme values as multiple electrons can contribute to the emission simultaneously and the encounter region is extended, \peijin{ consequently, the sources from counterstreaming electrons are more likely to be extended instead of point-like sources. as shown in Figure \ref{fig:imB1}, \ref{fig:imB2}, the sources in B is more extended than A and C, which also supports that B is from counterstream electrons}.

In our study, the sources in group A and group C have point-like sources, and the frequency-location track is aligned for all individual herringbone structures. This fits the scenario of the single-direction electron beams. Their frequency drift rate also has the same sign within the same group. Group B has extended sources and a relatively complex spatial structure, and an exceptionally high frequency drift rate for B1. B1 and B2 are also spatially separated, so these properties better fit the countersteam electron beams theory.


\section{Conclusion}

In this work, we performed a detailed imaging spectroscopy analysis for herringbone bursts in a Type II radio burst. From the three herringbone structure groups, we find groups A and C have relatively simple imaging sources, the frequency drift rate has the same sign within the same group, and the frequency drift rate absolute values are similar to the type IIIs in the same frequency range. Group B has relatively complex spatial structures in imaging and observed multiple sources with extended structures in both B1 and B2, and B1 has an exceptionally large frequency drift rate.
The Observation suggests that groups A and C are more likely to originate from single-direction beam electrons, and group B is more likely to be generated from counterstreaming beams.

\section*{Acknowledgments}
P.Z., D.E.M., and A.K. acknowledge the University of Helsinki Three-year Grant. D.E.M acknowledges the Academy of Finland project `RadioCME' (grant number 333859). E.K.J.K. acknowledges the European Research Council (ERC) under the European Union's Horizon 2020 Research and Innovation Programme Project SolMAG 724391 and the Academy of Finland Project 310445.
The authors acknowledge the Finnish Computing Competence Infrastructure (FCCI) for supporting this project with computational and data storage resources and thank Discoverer Petascale Supercomputer (Sofia, Bulgaria) for the provided computing resources.  A.K is supported by an appointment to the NASA Postdoctoral Program at the the NASA Goddard Space Flight Center (GSFC).
LOFAR \citep{van2013lofar} is the Low Frequency Array designed and constructed by ASTRON. It has observing, data processing, and data storage facilities in several countries, which are owned by various parties (each with its own funding sources), and that are collectively operated by the ILT foundation under a joint scientific policy. The ILT resources have benefited from the following recent major funding sources: CNRS-INSU, Observatoire de Paris and Universite Orleans, France; BMBF, MIWF-NRW, MPG, Germany; Science Foundation Ireland (SFI), Department of Business, Enterprise and Innovation (DBEI), Ireland; NWO, The Netherlands; The Science and Technology Facilities Council, UK; The Ministry of Science and Higher Education, Poland.

\bibliographystyle{aa}
\bibliography{typeIIHB}

\onecolumn
\appendix

\section{Imaging}

Interferometric imaging of the herringbone groups.

Group A:
\begin{figure}[H]
    \centering
    \includegraphics[width=0.8\textwidth]{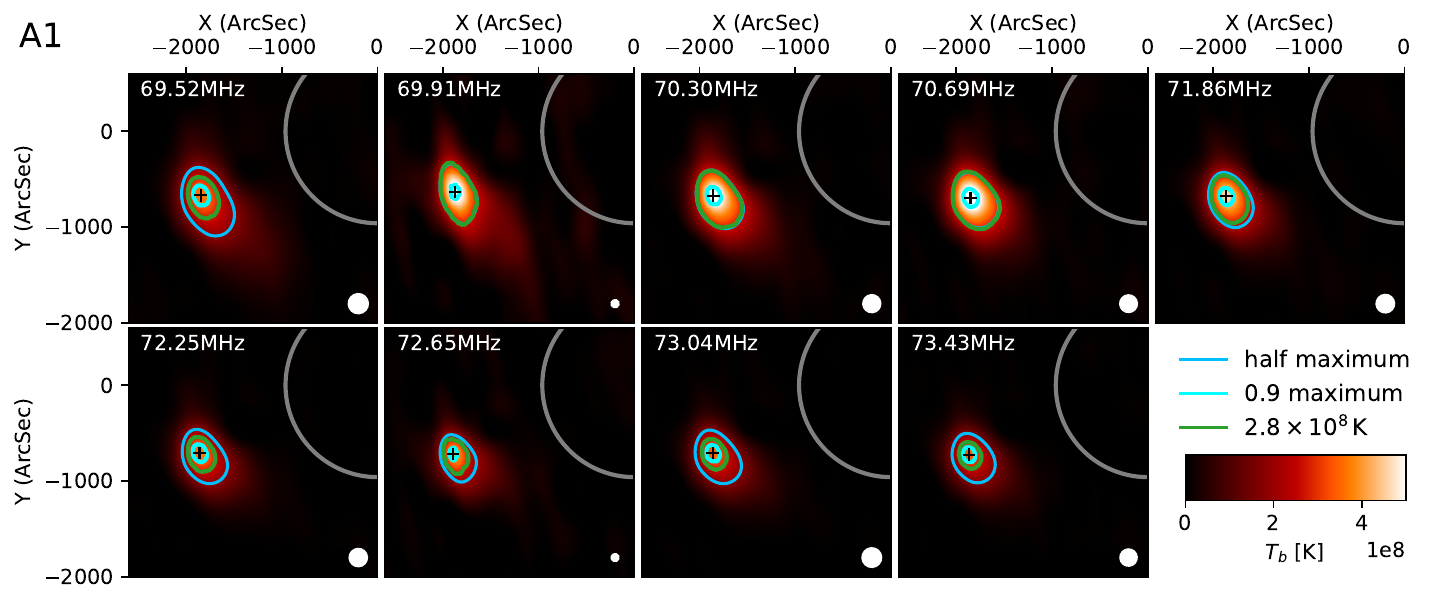}
    \caption{Interferometric imaging of the Herringbone fine structure A1, \peijin{the beam shape of the cleaned beam is marked as a white circle in the lower right corner of each panel}} 
    \label{fig:imA1}
\end{figure}

\begin{figure}[H]
    \centering
    \includegraphics[width=0.8\textwidth]{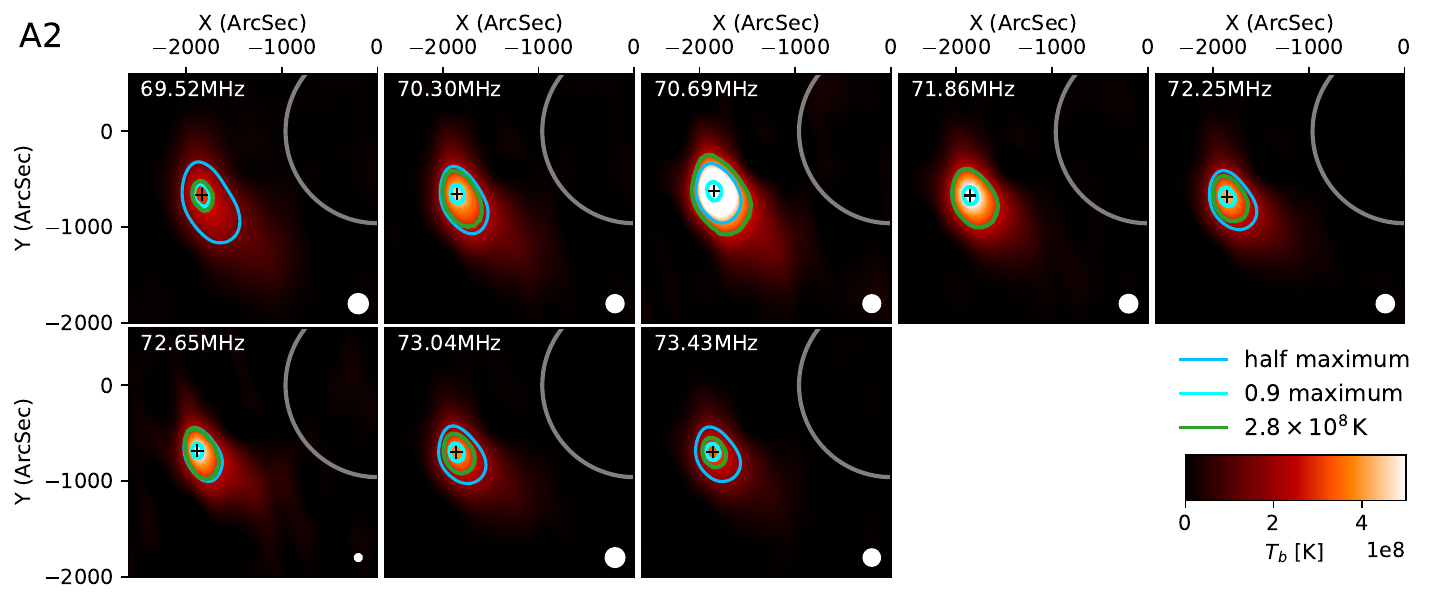}
    \caption{Interferometric imaging of the Herringbone fine structure A2, \peijin{the beam shape of the cleaned beam is marked as a white circle in the lower right corner of each panel}}
    \label{fig:imA2}
\end{figure}

\begin{figure}[H]
    \centering
    \includegraphics[width=0.8\textwidth]{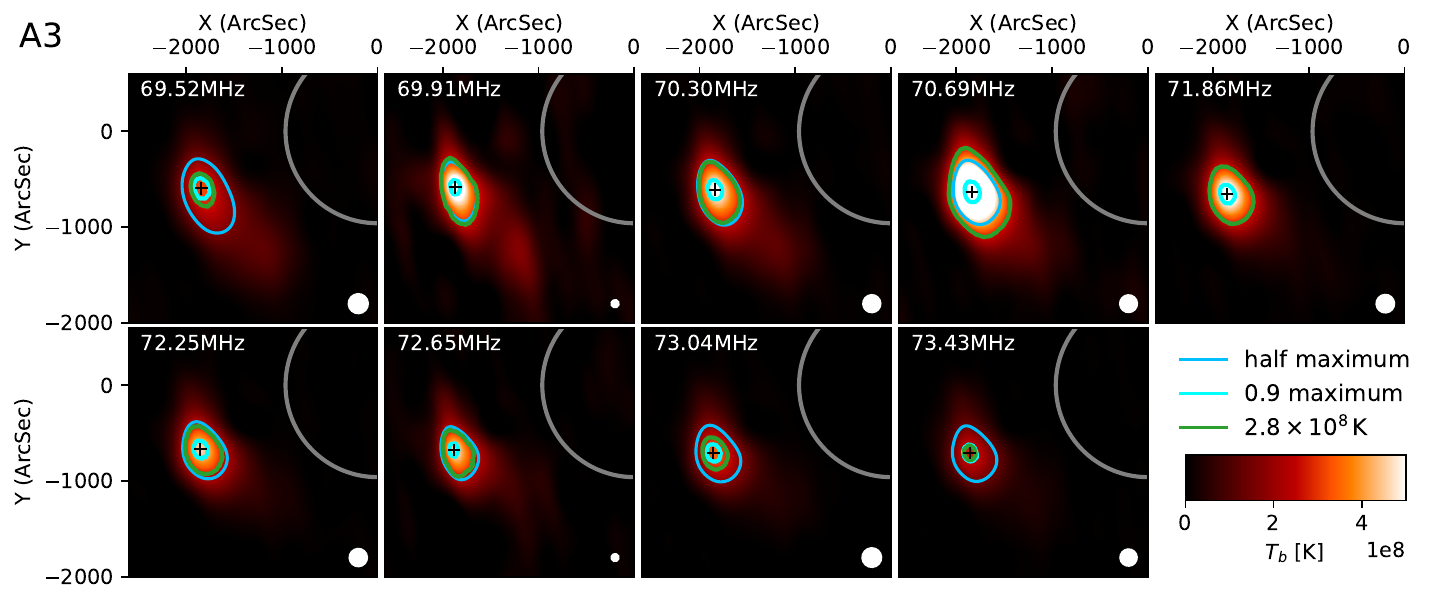}
    \caption{Interferometric imaging of the Herringbone fine structure A3, \peijin{the beam shape of the cleaned beam is marked as a white circle in the lower right corner of each panel}}
    \label{fig:imA3}
\end{figure}

\begin{figure}
    \centering
    \includegraphics[width=0.7\textwidth]{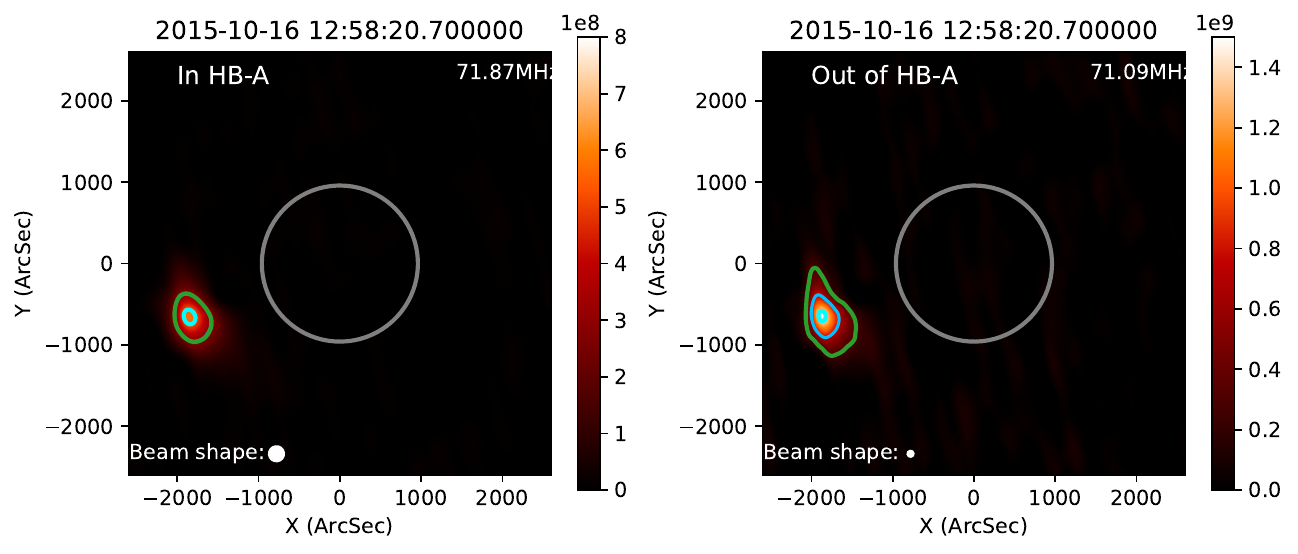}
    \caption{\peijin{Comparison of source location for Herringbone Group A and the overlapped structure, left panel is within the Herringbone, right panel is out of the Herringbone}}
    \label{fig:Ainout}
\end{figure}

Group B:
\begin{figure}[H]
    \centering
    \includegraphics[width=0.8\textwidth]{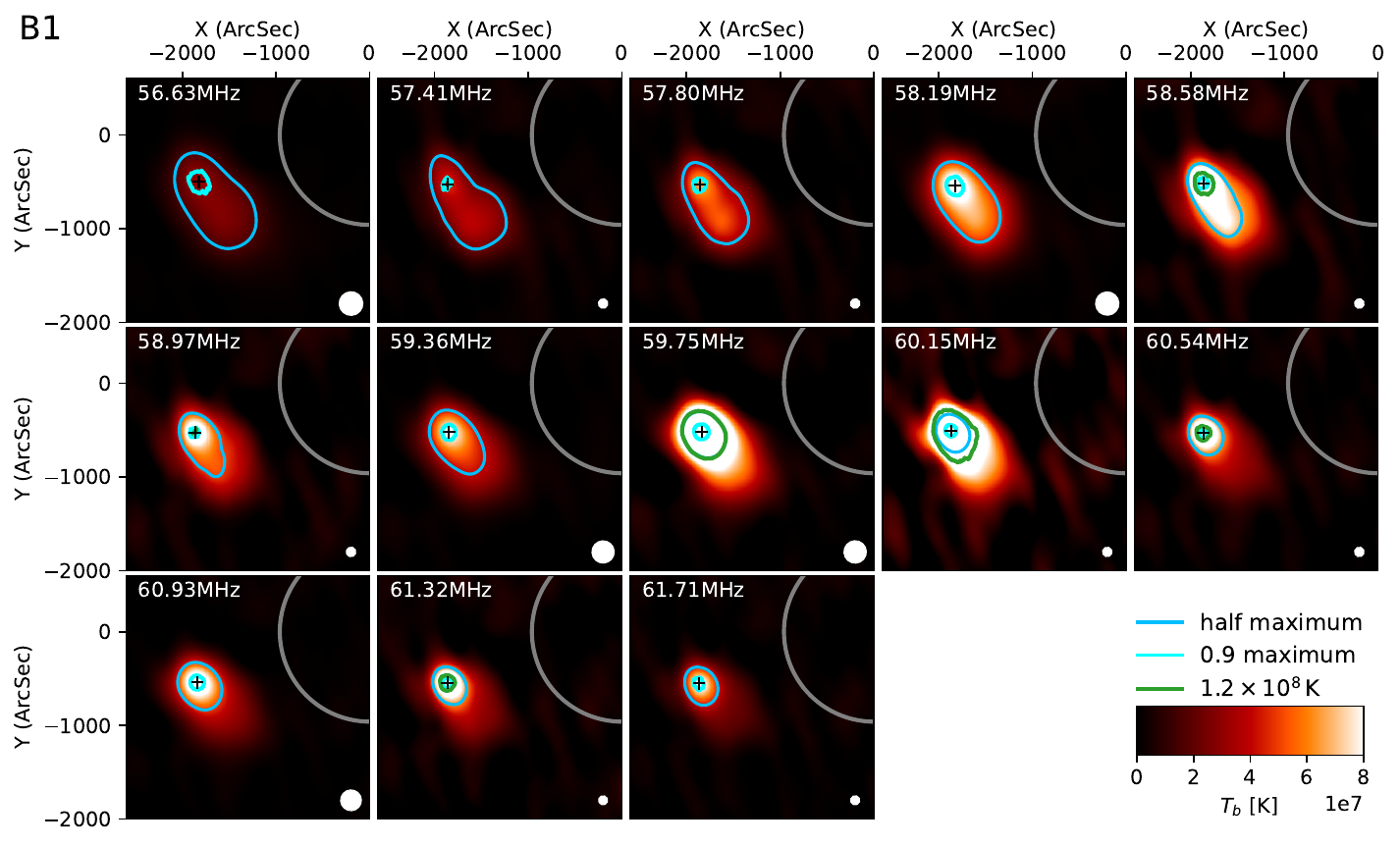}
    \caption{Interferometric imaging of the Herringbone fine structure B1, \peijin{the beam shape of the cleaned beam is marked as a white circle in the lower right corner of each panel}}
    \label{fig:imB1}
\end{figure}

\begin{figure}[H]
    \centering
    \includegraphics[width=0.8\textwidth]{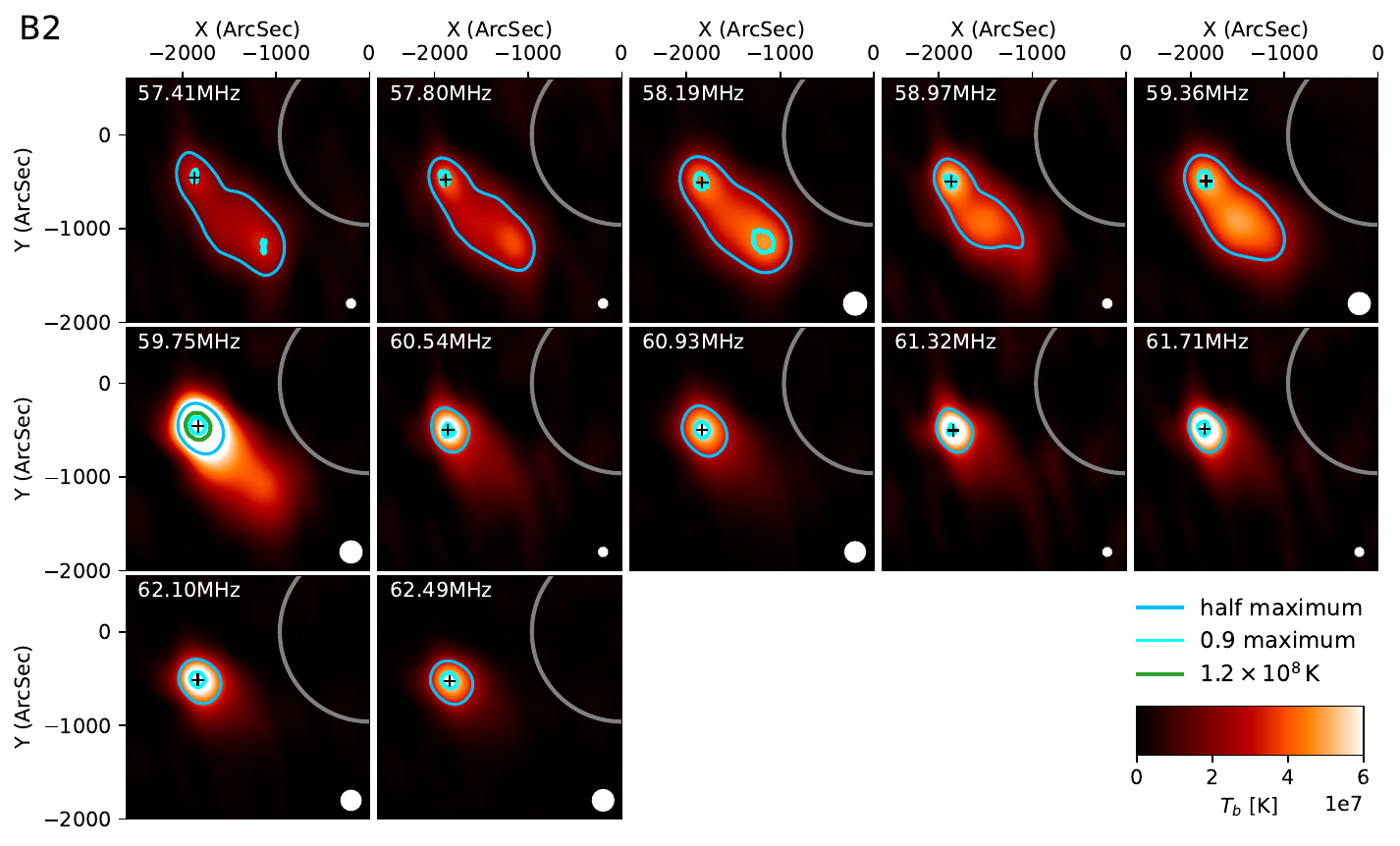}
    \caption{Interferometric imaging of the Herringbone fine structure B2, \peijin{the beam shape of the cleaned beam is marked as a white circle in the lower right corner of each panel}}
    \label{fig:imB2}
\end{figure}


\peijin{Adjacent time-freq point imaging of B2:}

\begin{figure}[H]
    \centering
    \includegraphics[width=0.7\textwidth]{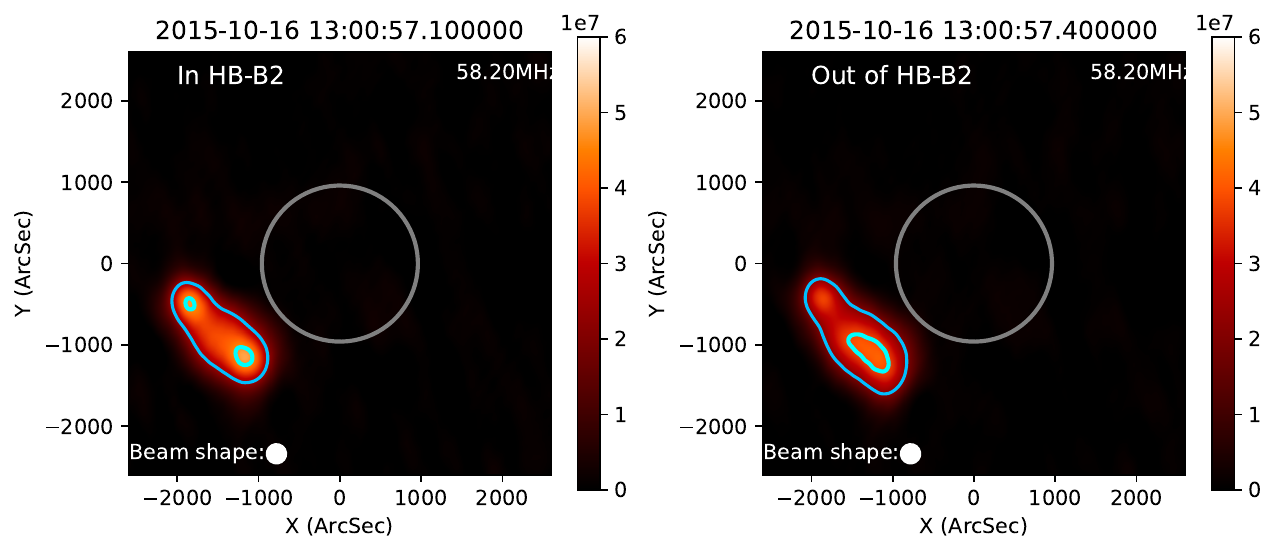}
    \caption{\peijin{Imaging comparison of in and out of the Herringbone structure B2.
    The left panel is inside the herringbone structure, the same as the 3rd panel in Fig \ref{fig:imB2}, the right panel is one time slot later than the left panel.}}
    \label{fig:inoutlabel}
\end{figure}

Group C:
\begin{figure}[H]
    \centering
    \includegraphics[width=0.6\paperwidth]{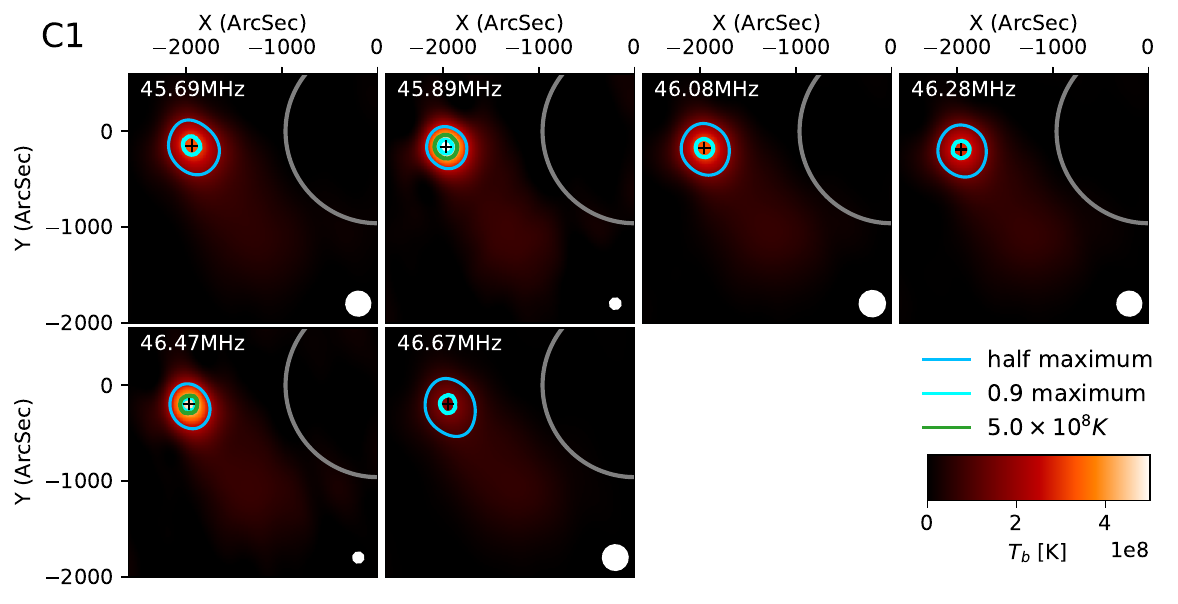}
    \caption{Interferometric imaging of the Herringbone fine structure C1, \peijin{the beam shape of the cleaned beam is marked as a white circle in the lower right corner of each panel}}
    \label{fig:imC1}
\end{figure}

\begin{figure}[H]
    \centering
    \includegraphics[width=0.6\paperwidth]{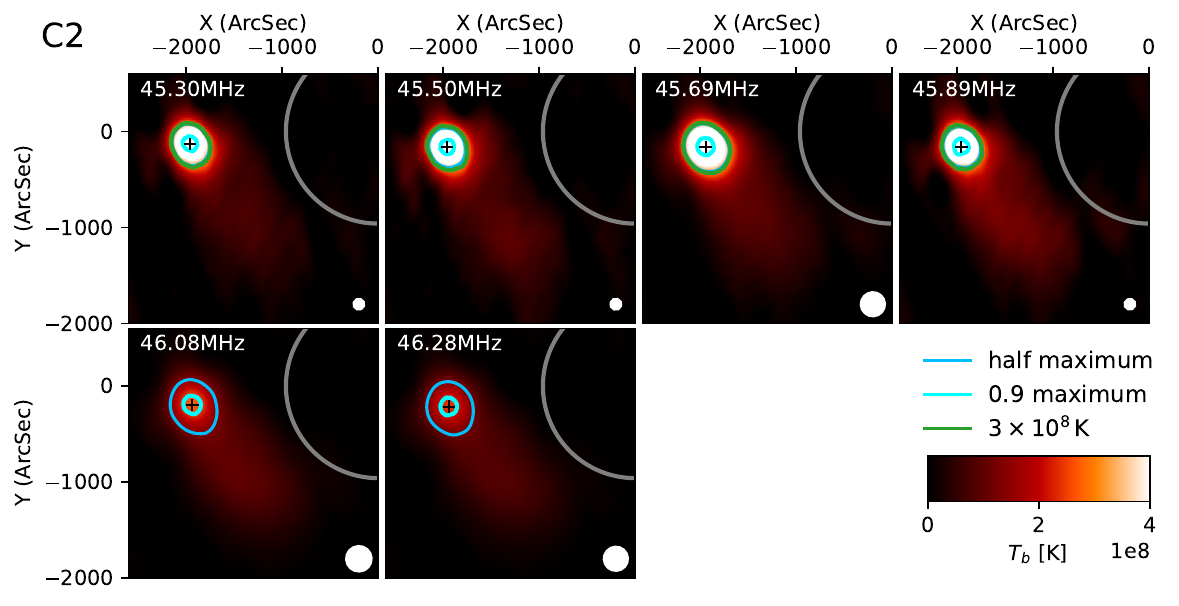}
    \caption{Interferometric imaging of the Herringbone fine structure C2, \peijin{the beam shape of the cleaned beam is marked as a white circle in the lower right corner of each panel}}
    \label{fig:imC2}
\end{figure}

\end{document}